\documentclass{article}

\usepackage{graphicx}
\usepackage{epsfig}

\title{
\bf{Novel Excitonic States in Quantum Hall Systems:
    Bound States of Spin Waves and a Valence Band Hole}
}

\author{
\large{John J. Quinn} \\ 
\small{University of Tennessee, Knoxville, Tennessee 37996, USA} \\
\large{Arkadiusz W\'ojs} \\
\small{University of Tennessee, Knoxville, Tennessee 37996, USA}, \\
\small{Wroclaw University of Technology, Wroclaw 50-370, Poland}
}

\begin{document}

\date{}

\maketitle

\begin{abstract}
If the Zeeman energy is small, the lowest energy excitations 
of a two dimensional electron gas at filling factor $\nu=1$
are spin waves (spin flip excitations).
At $\nu$ slightly larger (smaller) than unity, reversed spin
electrons (spin holes) can form bound states with $K$ spin waves 
that are known as skyrmions, $S_K^-$ (antiskyrmions, $S_K^+$).
It is suggested in this work that a valence hole can also bind 
$K$ spin waves to form an excitonic complex $X_K^+$, analogous 
to the $S_K^+$.
One spin hole of the $S_K^+$ is simply replaced by the valence 
hole.
At $\nu\le1$, a small number of $S_K^+$'s are present before 
introduction of the valence hole.
The $S_K^+$--$X_K^+$ repulsion leads to correlations and 
photoluminescence similar to those of a dilute 
electron--charged-exciton ($e$--$X^-$) system at $\nu\le{1\over3}$.
At $\nu\ge1$, the $S_K^-$--$X_K^+$ attraction can potentially 
lead to different behavior.\\
Keywords: {quantum Hall effect, skyrmion, photoluminescence}
\end{abstract}

\section{Introduction}

It has become clear \cite{bright,Yusa01} that neutral ($X$) and charged
($X^-$) excitons both play an important role in the photoluminescence
(PL) spectrum of realistic quantum Hall systems at high magnetic field
and low electron density (i.e. for filling factor $\nu\leq{1\over3}$).
This is true despite the ``hidden symmetry'' of the ideal theoretical
model (``ITM'' implies zero well width, $w$, and very high magnetic
field, $B$; impurity scattering will be ignored in all our calculations)
which suggests that PL occurs only from neutral exciton recombination
\cite{MacDonald90}.
At values of $\nu$ close to unity a considerable body of experimental
data exists \cite{Heiman88,Jiang98}, but no simple picture of the PL 
process has emerged.
In this note we suggest that positively charged excitonic complexes
($X_K^+$) consisting of $K$ spin waves (SW), each with angular momentum
$l_{\rm SW}=1$, bound to a valence hole ($v$) must occur for 
$\nu\approx1$, and that in real experimental systems at low temperature
these $X_K^+$ complexes could dominate the PL spectrum.
A SW consists of a reversed-spin-electron--spin-hole pair 
($e_{\rm R}h$) in the lowest Landau level (LL) of the conduction band.

Throughout this paper we contrast the predictions of the ITM with those
of realistic systems.
The latter requires the admixture of a number of LL's by the Coulomb 
interaction and taking into account the finite well width $w$.
Finite separation $d$ between the electron layer and the valence hole
layer can also be included.
These effects destroy the ``hidden symmetry'' which occurs when the
magnitude $|V_{ij}|$ of the Coulomb interaction is the same for any 
pair ($i,j$) selected from ($e_{\rm R}$, $h$, $v$).
The paper is organized in three main sections.
Section 2 contains a summary of the results predicted \cite{bright} 
for PL in dilute systems ($\nu\leq{1\over3})$.
Section 3 section contains a discussion of the elementary spin 
excitations \cite{skyrmion} of a system of $N$ electrons with $\nu$ 
close to unity in the absence of any valence band holes.
In Section 4 a valence hole is introduced into the $\nu\approx1$
system.
The formation of $X_K^+$ ($v+K\times{\rm SW}$) complexes is discussed 
using their analogy to skyrmions or antiskyrmions.
The implication for PL of the existence of a quantum liquid consisting
of electrons, skyrmions (antiskyrmions) and an $X_K^+$ for $\nu\geq1$
($\nu\leq1$) are discussed.
Some preliminary numerical results for simple realistic systems are 
presented.

\section{Energy Spectrum and PL for $\nu\leq{1\over3}$}

It has become rather standard to diagonalize numerically the Coulomb
interaction for a finite system of $N$ electrons confined to
a spherical surface which contains at its center a magnetic monopole
of strength $2Q$ flux quanta \cite{Haldane83}.
In the ITM only states of the lowest LL are included.
For realistic experimental systems (having a finite quantum well
width $w$ in a finite magnetic field $B$) both higher LL's and
the modification for the Coulomb matrix elements associated with the
envelope functions of the quantum well must be included.

\begin{figure}[tb]
\begin{center}
\begin{minipage}[t]{0.8\linewidth}
\includegraphics[width=\linewidth]{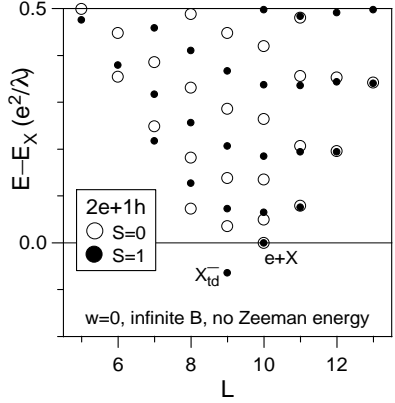}
\caption{
   The energy spectrum (energy $E$ vs.\ angular momentum $L$) 
   of the $2e$--$1v$ system on a Haldane sphere with the Landau 
   level degeneracy of $2Q+1=21$.
   $E_X$ is the exciton energy, and $\lambda$ is the magnetic 
   length.
   Different symbols distinguish between singlet ($S=0$) and triplet
   ($S=1$) states.}
\label{fig1}
\end{minipage}
\end{center}
\end{figure}

In Fig.~\ref{fig1} we present the energy spectrum for simple system
consisting of two electrons and one valence band hole at $2Q=20$ 
evaluated in the ITM and excluding the Zeeman energy \cite{bright}.
The solid dots are triplet electron states (the total spin of the 
pair of electrons $S=1$); the open circles are singlets ($S=0$).
The state labeled $e+X$ at angular momentum $L=10$ is a 
``multiplicative state'' consisting of an unbound electron and 
a neutral exciton ($X$).
Notice that only one bound state (labeled $X_{\rm td}^-$) occurs.
It is at $L=9$ and is called the ``dark triplet'' because it is 
forbidden to decay radiatively.

\begin{figure}[tb]
\begin{center}
\begin{minipage}[t]{0.8\linewidth}
\includegraphics[width=\linewidth]{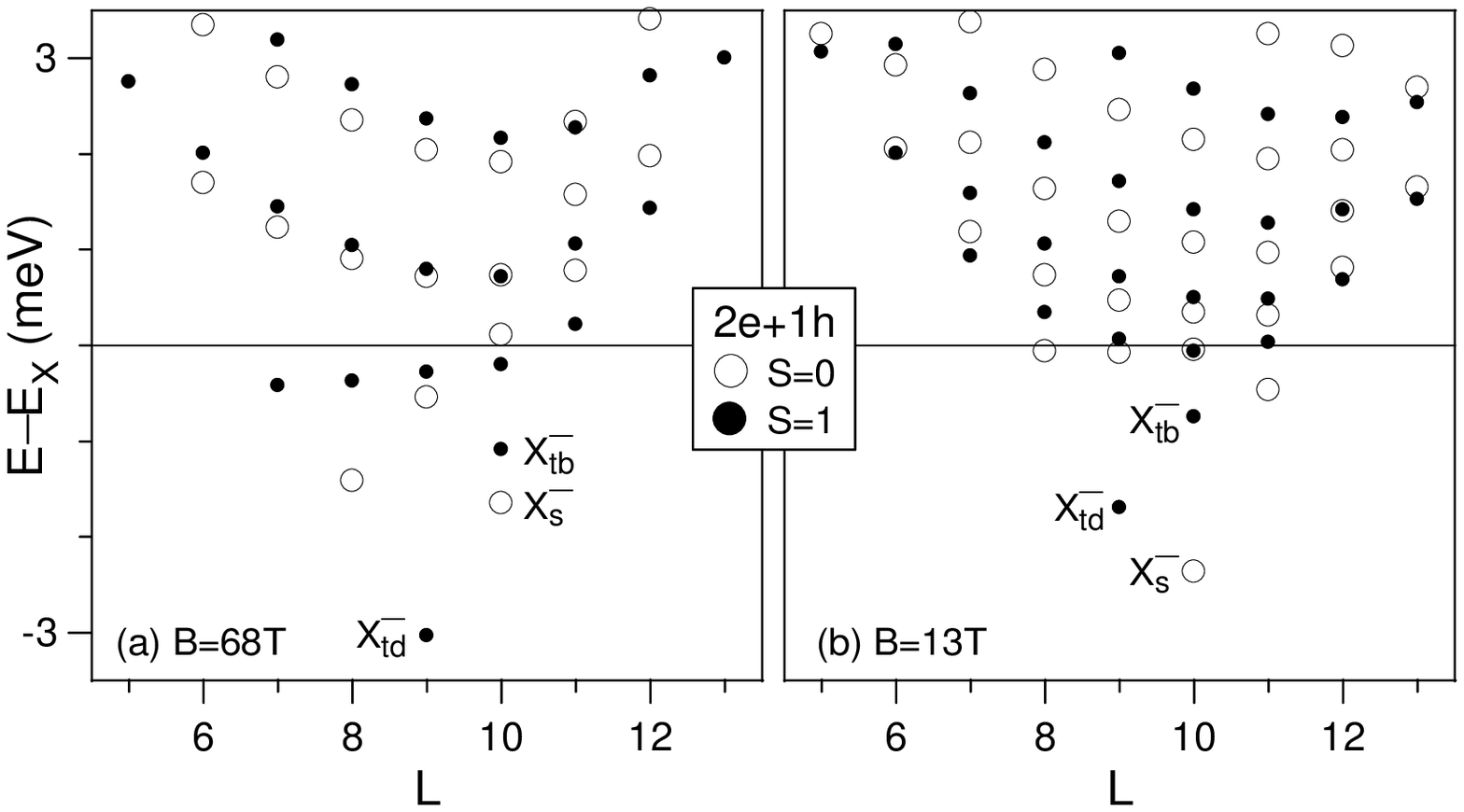}
\caption{
   Same as Fig.~\protect\ref{fig1}, but for a realistic GaAs 
   quantum well of width $w=11.5$~nm at the finite values of 
   $B$ as shown.
   The Zeeman energy has been included, and five LL's for both
   electrons and hole have been used in the diagonalization.}
\label{fig2}
\end{minipage}
\end{center}
\end{figure}

In Fig.~\ref{fig2} similar results are presented for a realistic system
consisting of a symmetric GaAs quantum well of width $w=11.5$~nm at
the finite values of the magnetic field $B=68$ and 13~T.
The appropriate electron Zeeman splitting has been included,
and only the lowest state of each triplet is shown.
To achieve even qualitative agreement with experimental data,
it has also been necessary to include a number of higher LL's,
particularly at the lower magnetic fields.
Five LL's were needed to obtain convergence in our calculations.
In Fig.~\ref{fig2}a, at the high magnetic field of 68~T, the 
$X_{\rm td}^-$ at $L=9$ is still the ground state, but singlet and 
another triplet bound states occur at other values of $L$
(the singlet $X_{\rm s}^-$ at $L=8$ and the bright triplet 
$X_{\rm td}^-$ at $L=10$ have roughly half the binding energy 
of the $X_{\rm td}^-$).
At $B=13$~T, as shown in Fig.~\ref{fig2}b, $X_{\rm s}^-$ is the ground 
state, and the $X_{\rm td}^-$ at $L=9$ and $X_{tb}^-$ at $L=10$ are 
excited states.
The spectrum is quite sensitive to the experimental parameters.
The well width $w$ enters the Coulomb interaction \cite{bright} through
$V(r)=e^2/\sqrt{r^2+d^2}$, where $d$ is proportional to $w$.
The cyclotron frequencies $\omega_{ce}(B)$ and $\omega_{cv}(B)$ 
for the electrons and valence band hole, and the Zeeman energy 
$E_{\rm Z}(B)$, are taken from experiment, after Refs.~\cite{Cole97} 
and \cite{Snelling91}, respectively.
For the values of the parameters used in our calculations, the singlet 
and triplet ground states cross at a value of $B$ of the order of 30~T.
This is in agreement with the calculations of Whittaker and Shields
\cite{Whittaker97} who used a different numerical approach.
Because exact diagonalization gives the eigenfunctions as well as 
the eigenvalues, it is straightforward to evaluate matrix elements 
of the luminescence operator $\hat L=\int\,{\rm d}^2r\,\hat\Psi_e(r)
\hat\Psi_v(r)$ between an initial state $\Phi_i$ of $N$ electrons and 
one valence hole, and final states $\Phi_f$ containing $N-1$ electrons.
$\hat\Psi_e$ and $\hat\Psi_v$ are the annihilation operators for 
an electron and valence hole respectively.
The oscillator strength for the transition \cite{Chen93} from 
$\left|\Phi_i\right>$ to $\left|\Phi_f\right>$ is proportional to 
$|\left<\right.\!\Phi_f|\hat L|\Phi_i\!\left.\right>|^2$.
For an isolated $X^-$ (where $N=2$) angular momentum conservation 
forbids the lowest triplet ($X_{\rm td}^-$) from decaying radiatively; 
the subscript ``d'' stands for ``dark''.
The $X_{\rm s}^-$ and $X_{\rm tb}^-$ have finite oscillator strengths 
which are of the same order of magnitude.

When additional electrons are present ($N>2$) radiative decay of the
$X_{\rm td}^-$ is not strictly forbidden, since in the recombination 
process an unbound electron can scatter, changing the momentum of the 
final state.
However, it was found that for $\nu\leq{1\over3}$ such decays are weak
because Laughlin correlations of the $X^-$ with unbound electrons
inhibit close collisions.
The amplitude for radiative decay of the $X_{\rm td}^-$ is estimated 
\cite{bright} to be smaller by one or more orders of magnitude than 
those of the $X_{\rm s}^-$ and $X_{\rm tb}^-$.
It was suggested in \cite{bright} that the $X_{\rm td}^-$ would be 
difficult to see in PL, and that the non-crossing peaks observed by 
Hayne {\sl et al.} \cite{Hayne99} were the $X_{\rm s}^-$ and 
$X_{\rm tb}^-$.
The presence of impurities relaxes the $\Delta L=0$ selection rule,
and the $X_{\rm td}^-$ peak is clearly observed at very low temperature
where the excited $X_{\rm tb}^-$ and $X_{\rm s}^-$ states are sparsely 
populated \cite{Yusa01}.
The agreement of experiment \cite{Yusa01} and the numerical predictions 
\cite{bright} reinforce the hope of using PL to understand correlations 
in fractional quantum Hall systems.

\section{Spin Excitations Near $\nu=1$}

\begin{figure}[tb]
\begin{center}
\begin{minipage}[t]{0.8\linewidth}
\includegraphics[width=\linewidth]{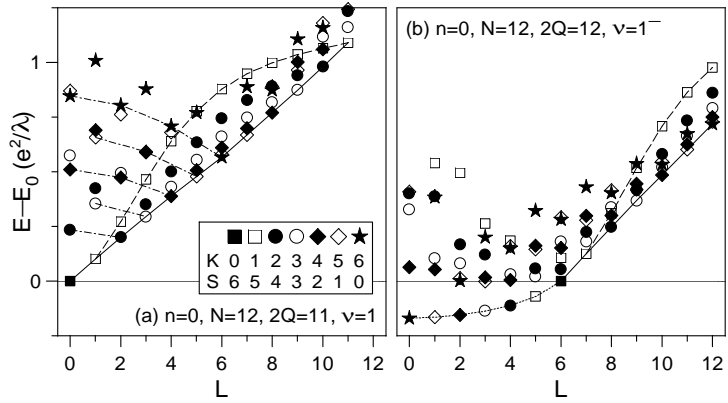}
\caption{
   The energy spectra of 12 electrons in the lowest LL
   calculated on Haldane sphere with $2Q=11$ (a) and 12 (b).}
\label{fig3}
\end{minipage}
\end{center}
\end{figure}

For filling factor $\nu$ equal to unity, the lowest energy excitations
are spin flip excitations which create a reversed spin electron,
$e_{\rm R}$, in the same $n=0$ LL leaving behind a spin hole, $h$, 
in the otherwise filled $\nu=1$ state.
Even when the Zeeman energy $E_{\rm Z}$ is zero, the Coulomb exchange 
energy will spontaneously break the spin ($\uparrow,\downarrow$) symmetry 
giving a spin polarized ground state.
In Fig.~\ref{fig3}a we show the low lying spin excitations of the
$\nu=1$ state (with $E_{\rm Z}$ taken to be zero) for a system of
$N=12$ electrons \cite{skyrmion}.
The solid square at $L=0$ is the spin polarized $\nu=1$ ground state 
with spin $S=6$.
The symbol $K={1\over2}N-S$ is the number of spin flips away from
the fully spin polarized state.
The band of open squares connected by a dashed line gives the spin 
wave dispersion $\varepsilon_{\rm SW}(L)$.
The angular momentum $L$ is related to wave vector $k$ by $L=kR$,
where $R$ is the radius of the spherical surface to which the $N$ 
electrons are confined.
The SW consists of a single $e_{\rm R}h$ pair; its dispersion can 
be evaluated analytically \cite{Kallin84}.
The solid circles, open circles, etc. represent states containing
2, 3, \dots\ spin flips (i.e. 2, 3, \dots\ $e_{\rm R}h$ pairs).
Dot-dashed lines connect low lying states with equal numbers of spin
flips.
It is interesting to note the almost straight line connecting the 
lowest energy states at $0\leq L\leq6$.
This can be interpreted as band of $K$ SW's each with $l_{\rm SW}=1$ 
with $L=K$.
The near linearity suggests that these $K$ SW's are very nearly
non-interacting.
In Fig.~\ref{fig3}b we show the numerical results for the situation in 
which $2Q=12$, so that one vacancy must be present in the $\nu=1$ state.
The notation is the same as in Fig.~\ref{fig3}a.
Here the $S=0$ state appears at $L=6$.
This is simply the single spin hole of $l=Q=6$ ($Q$ is the angular 
momentum of the lowest LL or angular momentum shell).
What is most interesting in the figure is the band of low lying
states containing $K=0$, 1, 2, \dots\ SW's bound to the spin hole.
The energy decreases with increasing $K$, but the decrease is slower
than linear.
In Fig.~\ref{fig3} we have neglected the Zeeman energy (taken the $g$-value
equal to zero).
For a finite $g$-value the Zeeman energy is simply $KE_{\rm Z}$, where
$E_{\rm Z}$ is the Zeeman energy of a single spin flip.
The Coulomb energy of the lowest state containing $K$ SW's is
$E_{\rm C}(K)\approx E_{\rm C}({1\over2}N)+\beta S^2$, where 
$E({1\over2}N)$ is the energy of the lowest $L=S=0$ state in 
Fig.~\ref{fig3}a, and $S$, the total spin, is equal to ${1\over2}N-K$.
Adding the Zeeman energy $KE_{\rm Z}$ leads to a total energy
$E(K)=E_{\rm C}({1\over2}N)+\beta({1\over2}N-K)^2+E_{\rm Z}K$.
This energy has a minimum at $K=K_0={1\over2}(N-E_{\rm Z}/\beta)$ 
implying that the lowest state contains approximately $K_0$ spin 
flips.
For $E_{\rm Z}=0$, $K_0={1\over2}N$, and the ground state is completely 
depolarized (i.e. $S=0$).
As $E_{\rm Z}$ is increased, the number of spin flips, $K$, in the 
lowest energy state decreases until at $E_{\rm Z}>\beta N$ only the 
spin hole in the $\nu=1$ state remains.
The state with the integral value of $K$ (closest to $K_0$) which gives 
the lowest energy is a measure of the size of the antiskyrmion, the state 
consisting of $K$ SW'ss bound to a spin hole in the $\nu=1$ state 
\cite{Sondhi93}.
By electron--hole symmetry the state containing one reversed spin 
electron, $e_{\rm R}$, in addition to the filled $\nu=1$ level will 
form an analogous skyrmion state consisting of $K$ SW'ss bound 
to the original $e_{\rm R}$.

The most stable skyrmion or antiskyrmion size depends weakly on the 
quantum well width for the $\nu\approx1$ state, but for $\nu\approx3$, 
5, \dots\ the well width $w$ must be of the order of a few times the 
magnetic length in order to obtain stable bound states of SW's and 
spin holes or reversed spin electrons \cite{skyrmion,Cooper97}.
As reported by Melik-Alaverdian {\sl et al.} \cite{Melik99}, the 
inclusion of the admixture of higher LL's caused by the Coulomb 
interaction weakly affects the skyrmion energy spectrum, particularly
when the finite wel width $w$ is also taken into account.

The skyrmion and antiskyrmion states $S_K^\pm$ are quite analogous to 
the excitonic $X_K^\pm$ states of valence band holes interacting with 
conduction band electrons.
In the ITM, a valence hole has exactly the same interactions as 
a spin hole in the $\nu=1$ state of the conduction band.
In fact these two types of holes can probably be distinguished by an
isospin as is done for electrons on different layers of a bilayer
system \cite{DasSarma94}.
The spectrum and possible condensed states of a multicomponent Fermion
liquid containing electrons, $X_1^-$, $X_2^-$, \dots, etc., has been
considered by W\'ojs {\sl et al.} \cite{x-cf}.
Exactly the same ideas are applicable to a liquid of electrons and 
skyrmions or antiskyrmions of different sizes.
The only difference is that the skyrmion $S^-=he_{\rm R}^2$ is 
stable while the $X^-=ve^2$ has a finite lifetime for radiative 
recombination of an electron--valence-hole pair.

When there are $N_h$ spin holes in the $\nu=1$ level (or $N_e$ reversed 
spin electrons in addition to the filled $\nu=1$ level) and when $N_h$
(or $N_e$) is much smaller than $N\approx2Q+1$, the degeneracy of the 
filled lowest LL, then the most stable configuration will consist of 
$N_h$ antiskyrmions (or $N_e$ skyrmions) of the most stable size.
These antiskyrmions (or skyrmions) repel one another.
They are positively (or negatively) charged Fermions with standard LL
structure, so it is not surprising that they would form either a Wigner
lattice or a Laughlin condensed state with $\nu$ for the antiskyrmion
(or skyrmion) equal to an odd denominator fraction as discussed in 
Refs. \cite{skyrmion,x-fqhe,Brey95}.

\section{Photoluminescence Near $\nu=1$}

In the ITM, a valence hole acts exactly like a spin hole in the $\nu=1$ 
level of the conduction band.
Therefore we would expect an excitonic complex consisting of $K$ SW's
bound to the valence hole to be the lowest energy state, in the same 
way that the antiskyrmion consisting of $K$ SW's bound to a spin 
hole in the $\nu=1$ level gives the lowest energy state when $E_{\rm Z}$
is less than $\beta N$.
For a small number of valence holes, the $X_K^+=v(e_{\rm R}h)^K$ 
excitonic complexes formed by each valence hole will repel one another.
If a small number of antiskyrmions are already present (for $\nu<1$), 
the positively charged antiskyrmion--charged-exciton repulsion will 
lead to Laughlin correlations or Wigner crystallization of the 
multicomponent Fermion liquid.
Just as for the $X^-$ excitons in the dilute regime, the PL at low
temperature will be dominated by the $X_K^+\rightarrow S_{K'}^++\gamma$ 
process, with $K'=K$ or $K-1$ depending on spin of the annihilated
valence hole (i.e. on the circular polarization of the emitted photon 
$\gamma$).
This corresponds to the most stable $X_K^+$ undergoing radiative 
$ev$ or $e_{\rm R}v$ recombination and leaving behind an antiskyrmion 
consisting of $K$ or $K-1$ SW's bound to a spin hole of the $\nu=1$ 
state.
Because the valence hole and the spin hole in the $\nu=1$ conduction 
level are distinguishable (or have different isospin) even in the 
ITM this PL is not forbidden.
It will be very interesting to see how realistic sample effects 
(finite well width, LL admixture, finite separation between the 
electron and valence hole layers) alter the conclusions of the ITM.

\begin{figure}[tb]
\begin{center}
\begin{minipage}[t]{0.8\linewidth}
\includegraphics[width=\linewidth]{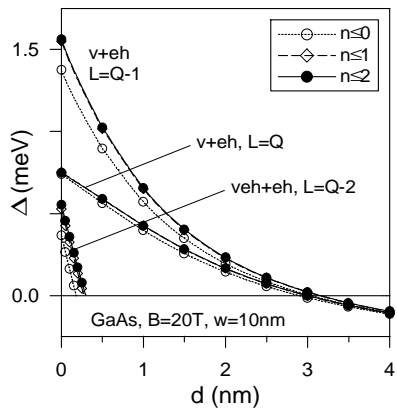}
\caption{
   Binding energies of $X_1^+=v(e_{\rm R}h)$ at $L=Q-1$ and $Q$, 
   and of $X_2^+=v(e_{\rm R}h)^2$ at $L=Q-2$ 
   as a function of $d$, the $e$--$v$ layer separation.
   The calculations are for a GaAs quantum well of width
   10~nm at a magnetic field $B=20$~T.
   Different curves include one, two, and three LL's 
   for the valence hole.}
\label{fig4}
\end{minipage}
\end{center}
\end{figure}

For $\nu\geq1$, negatively charged skyrmions are present before the
introduction of the valence holes.
The skyrmions are attracted by the $X_K^+$ charge exciton, but 
how this interaction affects the PL can only be guessed.
We are currently investigating real sample effects in systems 
containing a small number of skyrmions (or antiskyrmions) and 
valence band excitonic complexes.
As one preliminary example we show in Fig.~\ref{fig4} the binding 
energy of the $X_1^+=v(e_{\rm R}h)$ and $X_2^+=v(e_{\rm R}h)^2$ 
complexes for different values of the total angular momentum $L$ 
as a function of the separation between the electron and valence 
hole layers.
The calculation was done for parameters corresponding to a GaAs 
quantum well of width $w=10$~nm, at a magnetic field of 20~T.
Different symbols (open circles, open diamonds, and solid circles) 
are for calculations in which one, two, or three LL's for the
valence hole have been included (inter-LL excitations of conduction
electrons are less important due to their smaller effective mass.
Clearly, binding energies decrease with increasing layer separation 
as expected.

We believe that numerical diagonalization for realistic models 
including LL admixture and finite well width should explain the 
behavior of PL for electron filling factors close to unity.
The qualitative behavior expected has been discussed in this note.
Realistic ``numerical experiments'' are being carried out to check
whether the expected behavior is correct.
These results will be reported elsewhere.

\section*{Acknowledgment}

The authors wish to acknowledge partial support from the Materials 
Research Program of Basic Energy Sciences, US Department of Energy
and thank I.~Szlufarska for helpful discussions.
AW acknowledges partial support from the Polish State Committee 
for Scientific Research (KBN) grant 2P03B05518.

\end{document}